\documentstyle[12pt,a4,subeqn,subeqnarray]{article}

\catcode `@=11 \@addtoreset{equation}{section} \catcode `@=12
\begin{document}
\input epsf
\def\be{\begin{equation}}
\def\ee{\end{equation}}
\def\ba{\begin{array}{l}}
\def\ea{\end{array}}
\def\beq{\begin{eqnarray}}
\def\eeq{\end{eqnarray}}
\def\eq#1{(\ref{#1})}
\def\del{\partial}
\def\A{{\cal A}}

\renewcommand\arraystretch{1.5}

\begin{flushright}
hep-th/9705215\\
TIFR-TH-97/09\\
May 1997
\end{flushright}
\begin{center}
\vspace{3 ex}
{\large\bf THE EMERGENCE OF SPACE-TIME GRAVITATIONAL PHYSICS AS AN
EFFECTIVE THEORY FROM THE $c=1$ MATRIX MODEL}\\
\vspace{8 ex}
Avinash Dhar$^\dagger$ \\
{\sl Theoretical Physics Group, Tata Institute of Fundamental Research,} \\
{\sl Homi Bhabha Road, Mumbai 400 005, INDIA.} \\
\vspace{15 ex}
\bf ABSTRACT\\
\end{center}
\vspace{2 ex}
We discuss further a recent space-time interpretation of the 
$c=1$ matrix model which retains both sides of the inverted harmonic
oscillator potential in the underlying free fermion theory and  
reproduces the physics of the discrete state moduli of two-dimensional
string theory. We show that within this framework the linear
tachyon background in flat space arises from the fermi vacuum.
We argue that this framework does not suffer from any obvious 
nonperturbative inconsistency. We also identify and
discuss a class of nearly static configurations in the free fermion
theory which are interpreted as static metric backgrounds in space-time.
These backgrounds are classically absorbing --- a beam of tachyons
thrown at such a background is only partly reflected back --- and are
tentatively identified with the eternal back-hole of 2-dimensional
string theory.

\vfill

\noindent\vrule height 0 pt depth 0.4 pt width 2in

\noindent$^\dagger$e-mail: adhar@theory.tifr.res.in

\clearpage

\section{Introduction}

One of the most difficult aspects of the $c=1$ matrix model has been
its connection with the space-time physics of 2-dimensional string
theory\footnote{For a brief review and a list of original references
see ref. [1]}. In particular, there is no hint in the matrix model of an
interpretation as a theory of gravity. Indeed, it is even possible to
argue that gravitational effects are absent in this model! However,
a careful analysis of this model and its comparison with known facts
from perturbative 2-dimensional string theory has revealed a
different story. Space-time gravitational physics is now understood to
be encoded in the $c=1$ matrix model in a subtle and unexpected way and
2-dimensional gravity emerges from it as a low-energy effective theory! 

There are two key observations that have made this possible: (i) In
the double scaling limit [2] the $c=1$ matrix model is equivalent to a
theory of nonrelativistic, noninteracting fermions in an inverted
harmonic oscillator potential in one space dimension [3]. The
semiclassical physics of the matrix model is, therefore, described by a
fermi liquid theory. Small fluctuation of the fermi surface around the
fermi vacuum are described by a massless scalar excitation [4--6]. The
scattering amplitudes for this massless particle are not identical to
the tachyon scattering amplitudes of 2-dimensional string theory, but
are related to these by momentum-dependent `leg-pole' factors [7--9].
Although this factor is a pure phase in momentum space, in position
space it relates the wavefunction of the tachyon of 2-dimensional string
theory to the wavefunction of the massless excitation of the matrix model 
by a nonlocal  transformation. 
It is this nonlocal transformation
which gives rise to all of space-time gravitational physics of the
string theory, which is otherwise absent in the matrix model [10]. 
(ii) The
symmetric inverted harmonic oscillator potential has two essentially
decoupled sides in the semiclassical limit. The abovementioned
identification of a massless scalar excitation and its space-time
interpretation is based on small fluctuations of the fermi surface on
any one side of the potential only (Fig. 1).
\begin{figure}[htb]
\centerline{\epsfxsize=5.2in\epsfbox{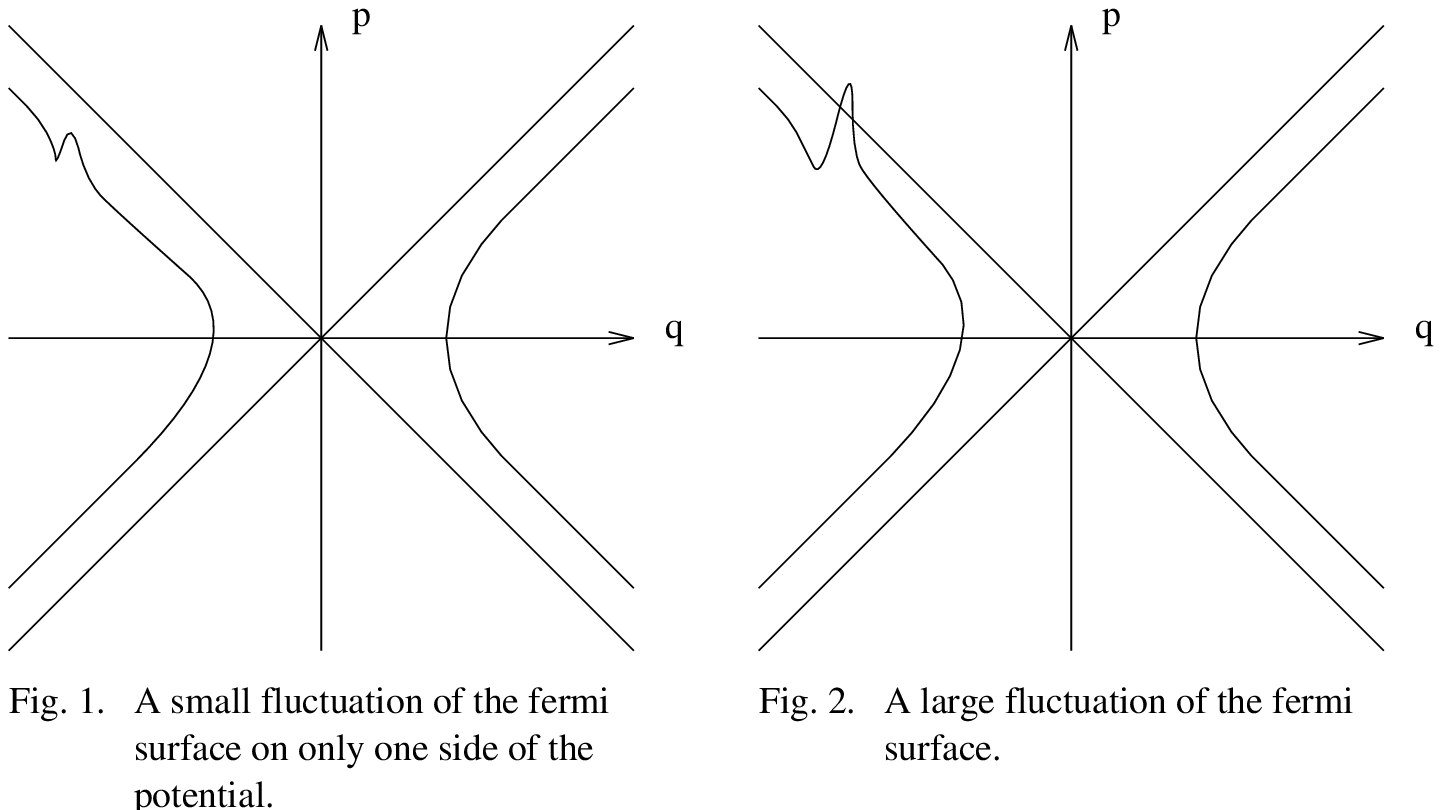}}
\end{figure}
This would seem to be a
sensible thing to do because in the semiclassical limit tunnelling to
the other side can be ignored and because ``small'' classical
fluctuations of the fermi surface cannot go over to the other side
(unlike the ``large'' ones which cross the asymptotes, (Fig. 2)).
The issue of the other side of the potential is then postponed to a
discussion of nonperturbative effects. 

This argument, however, cannot
be right, if the nonrelativistic fermion theory is the microscopic
theory underlying string theory, for the following reason. String
theory is a theory of gravity and so for consistency the space-time
metric must couple to the energy-momentum tensor of the theory.
In particular, it must couple to the total energy of any field
configuration in space-time. Now, if the matrix model is the
microscopic theory underlying 2-dimensional string theory then it
should be possible to map
any field configuration in space-time to some field configuration in
the free fermion theory. Therefore, the energy of this space-time
configuration may be computed using the microscopic Hamiltonian. 
If we decide to retain both sides of the inverted harmonic oscillator
potential, then a generic fluctuation will consist of fluctuations of
the fermi surface on both sides of the potential (Fig. 3)
and so the
other side will contribute to the total energy, even in the
small-field semiclassical limit when the two sides are otherwise 
decoupled in the matrix model. 
\begin{figure}[hbt]
\centerline{\epsfxsize=2.8in\epsfbox{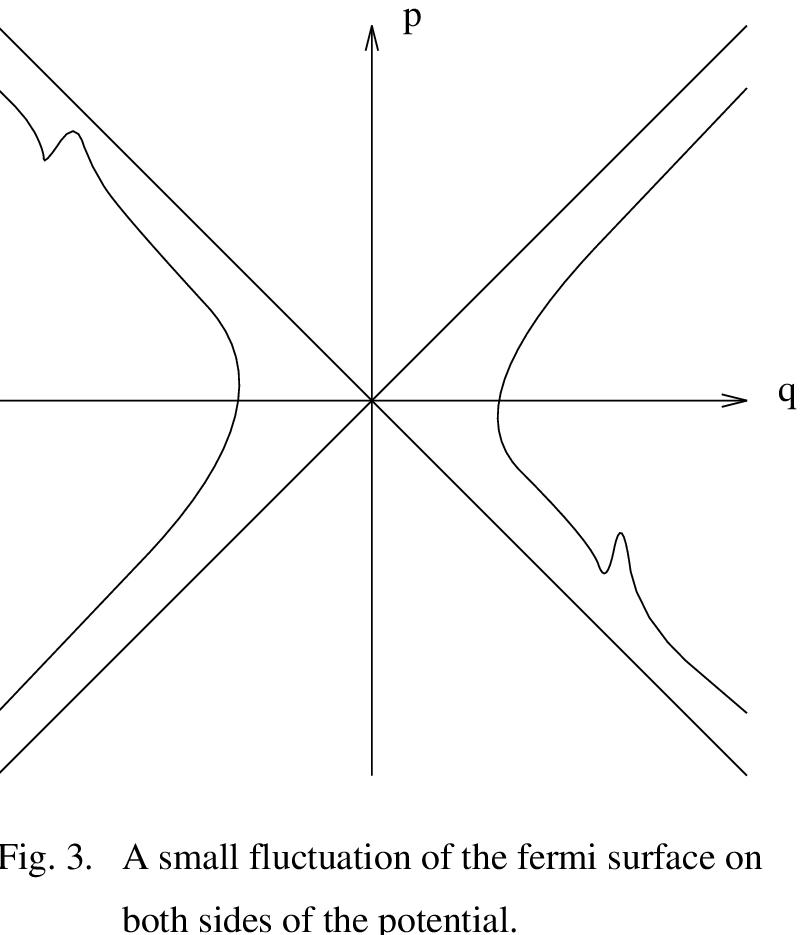}}
\end{figure}

In
the interpretation in (i), based on one side of the potential only [10],
the space-time metric is found to couple only to the energy of fermi
fluctuation on that side of the potential, even in the generic case
when there is a fluctuation of the fermi surface on the other side
also. This is clearly inconsistent with gravitational physics, unless
we decide to remove the other side of the potential from the start by
introducing a wall (Fig. 4).
\begin{figure}[htb]
\centerline{\epsfxsize=5in\epsfbox{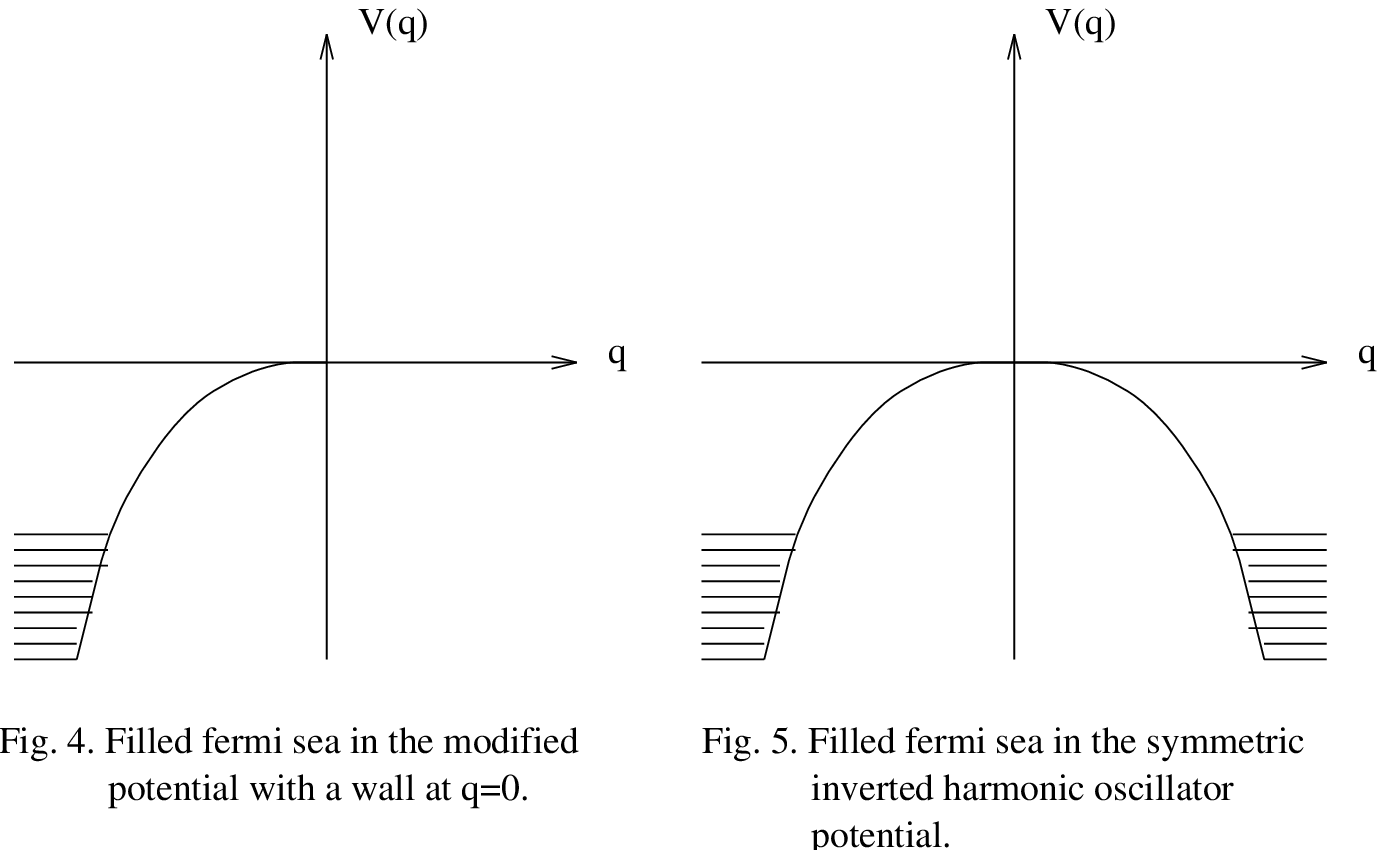}}
\end{figure}

Thus, to arrive at a consistent space-time interpretation of this
model, we must decide on the fate of the other side of the potential
already in the small-field semiclassical limit. The interesting case is
the one in which both sides of the potential are retained (Fig. 5)
since it is this case that describes the discrete state moduli of
2-dimensional string theory [11]. This possibility arises because in this
case at perturbation theory level we have {\it two} massless
scalar fields in the matrix model. One combination of these maps to
the tachyon of string theory while the other combination encodes the
discrete state moduli.

The organization of this paper is as follows. In the next section we
will discuss in some detail the mapping of the matrix model to
space-time outlined in (ii) above. We will identify and give the
precise relation of the parameters characterising background tachyon
and metric perturbations of 2-dimensional string theory with the
fields describing the fluctuations of the fermi surface. This is
essentially a review of [11]. In Sec. 3 we will show that the linear
tachyon background of string theory in flat space originates from the
fermi vacuum of the microscopic theory. Together with the
results of Sec. 2 this then implies that there is a mapping of
{\it all} the states of the fermi theory to space-time
configurations of 2-dimensional string theory, at least in
perturbation theory. In Sec. 4 we will identify some configurations in
the fermi theory which correspond to essentially static metric
backgrounds in their space-time interpretation. The ``mass'' parameter
which characterises these metric backgrounds is continuous. We will
further show that these configurations absorb a part of incident
tachyon flux. This makes a tentative identification of these
configurations with the eternal black-hole of 2-dimensional string
theory very plausible. In Sec. 5 we will discuss a potential
nonperturbative inconsistency pointed out by Polchinski [12] and
argue that the identification of the weak field expansion parameter in
[12], which is one of the crucial elements in Polchinski's argument,
may not be correct. Finally we will conclude with some closing 
remarks in Sec. 6.

\vskip 2em

\section{Discrete-state moduli from the free Fermi theory}

The nonlocal transformation of the wavefunctions of asymptotic ``in''
and ``out'' states of the matrix model, which describes perturbative
gravitational physics expected in 2-dimensional string theory, may be
expressed in terms of the fermi surface fluctuations as follows:
\beq
{\cal J}_{\rm in} (x^+) &=& \int^{+\infty}_{-\infty} d \tau \ f \left(
\left| {\mu \over 2}\right|^{1/2} \ e^{\tau - x^+}\right) \bar\eta_{+
\rm in} (\tau) , \\ [3mm]
{\cal J}_{\rm out} (x^-) &=& \int^{+\infty}_{-\infty} d \tau \ f \left(
\left| {\mu \over 2}\right|^{1/2} \ e^{-\tau + x^-}\right) \bar\eta_{-
\rm out} (\tau) .
\eeq
where $x^\pm \equiv t \pm x , (x,t)$ being space-time labels; ``in''
and ``out'' refer to $t \rightarrow - \infty$ and $t \rightarrow +
\infty$ limits respectively. The space coordinate $x$ is taken to be
large positive keeping $x^+$ fixed as $t \rightarrow - \infty$ and $x^-$
fixed as $t \rightarrow + \infty$. The function $f$ is given by
\be
f(\sigma) \equiv {1 \over 2 \sqrt{\pi}} J_0 \left(2
\left({2\over\pi}\right)^{1\over 8} \sqrt{\sigma}\right) , \sigma \leq
0
\ee
The fields $\bar\eta_{\pm}$ refer to the collective field
parametrization of the fluctuations of the fermi surface (relative to
the fermi vacuum) which is assumed to have a quadratic profile. For
more details on this and on our definitions and notations we refer the
reader to ref. [11]. 

There are two important assumptions that have been made in writing
eqns. (2.1) and (2.2). One of these we have already mentioned, namely
that these equations assume a quadratic profile for the fluctuations of
the fermi surface. This is clearly an undesirable restriction since in
general the fluctuations need not have a quadratic profile. Moreover,
it is also desirable to write these equations in a form that does not
refer to any specific parametrization of the fermi surface
since the physics of the tachyon
of 2-dimensional string theory should depend only on the profile of
the fluctuations and not on any specific parametrization of these.
Such a more general form of eqns. (2.1) and (2.2) has been discussed
in ref. [13]. Actually, in this reference a more general
transformation, which is valid even away from the asymptotic $t
\rightarrow \pm \infty$, $x \rightarrow + \infty$ regions, is obtained
from the requirement that the field ${\cal J} (x, t)$ which
this transform defines should satisfy the $\beta$-function equation of
motion of the tachyon field of 2-dimensional string theory. It turns
out that for the purposes of computing tachyon scattering amplitudes
only the asymptotic form of the transform is relevant, since the
corrections drop out at large positive values of $x$. This asymptotic form
of the nonlocal transformation is given by
\be
{\cal J}(x,t) = \int dp \ dq \ f (-q e^{-x}) \ \delta {\cal U} (p, q,
t) + O (e^{-2x})
\ee
where $\delta {\cal U}$ denotes a fluctuation of the fermi fluid
density ${\cal U} (p, q, t)$ in phase space $(p, q)$. We refer the
reader to [14] for a detailed account of the phase space formulation
of the double-scaled fermion theory and its connection with the
collective field theory. For the present purposes, however, the
summary given in [11] will be sufficient. We have used the notations
and definitions of this reference throughout the present work.

In the asymptotic space-time regions and for a quadratic profile of the
fluctuation $\delta {\cal U}$, the above definition of ${\cal J}
(x,t)$ reproduces eqns. (2.1) and (2.2). We stress, however, that this
definition of ${\cal J} (x,t)$ allows us to write ${\cal J}_{\rm in}$
and ${\cal J}_{\rm out}$ more generally for {\it any} profile of the phase
space density fluctuation. This definition is, in fact, even
independent of any parametrization used for the fluctuation, as we
shall see later.

The second assumption used in writing eqns. (2.1) and (2.2) is that the
fluctuation of the fermi surface on only one side of the potential is
relevant, even when there is no wall in the middle, in the
semiclassical limit. It is for this reason that the $q$ integration on
the right hand side of eqn. (2.4) is restricted to the region $-
\infty > q \geq - |2\mu|^{1/2}$. However, as we have argued in the
Introduction, in the theory without a wall (i.e. with a symmetrical
two-sided inverted harmonic oscillator potential) it is inconsistent
to ignore the ``other side'' of the potential even in the
semiclassical limit. We, therefore, need a further generalization of
eqn. (2.4) which includes fluctuations of the fermi surface on both
sides of the potential. The simplest possibility that suggests itself,
because of the symmetry of the potential, is a symmetrical transform.
Since the function $f$ is defined only for positive values of its
argument, and since $q$ has opposite signs on the two sides of the
potential, this symmetrical double-sided transform must take the form
\be
{\cal J}(x,t) = 2^{-1/2} \int dp \ dq \ f \left(2^{1/4} |q| e^{-x}
\right) \delta {\cal U} (p, q, t) + O (e^{-2x})
\ee
The overall factor of $2^{-1/2}$ and the extra factor of $2^{1/4}$ in
the argument of $f$ relative to the expression in eqn. (2.4) have been
put there only for later convenience. The $q$ integration in eqn. (2.5) now
extends over the entire real line and so this definition of ${\cal J}
(x,t)$ includes density fluctuations of the fermi fluid on both sides
of the potential.

The question we now ask is: Does the field ${\cal J} (x,t)$ defined in
eqn. (2.5) reproduce the tree-level scattering amplitudes of the
tachyon of 2-dimensional string theory? This question is easily
answered since eqn. (2.5) allows us to relate the asymptotic ``in''
and ``out'' wavefunctions of the field ${\cal J} (x,t)$. In this
context it is useful to note that, by a change of integration
variables and use of the equation of motion of ${\cal U}, \
(\partial_t + p \partial_q + q \partial_p) {\cal U} (p, q, t) = 0$,
the right hand side of eqn. (2.5) may be re-expressed in terms of
density fluctuation $\delta {\cal U}_0 (p,q) \equiv \delta {\cal U}
(p, q, t_0)$ at some initial time $t = t_0$:
\be
{\cal J} (x,t) = 2^{-1/2} \int \ dp \ dq \ f \left(2^{1/4} |Q (t-t_0)|
e^{-x}\right) \delta {\cal U}_0 (p, q) + O (e^{-x}) ,
\ee
where $Q(\tau) \equiv q \cosh \tau + p \sinh \tau$. Note that the
right hand side of eqn. (2.6) does not depend on $t_0$. This follows from
the equation of motion of ${\cal U}$. 

We may now write down expressions for ${\cal J}_{\rm in} (x^+)$
(obtained from eqn. (2.6) in the limit $t \rightarrow - \infty$,
keeping $x^+$ fixed) and ${\cal J}_{\rm out} (x^-) (t \rightarrow +
\infty, x^-$ fixed):
\beq
{\cal J}_{\rm in} (x^+) &=& 2^{-1/2} \int dp \ dq \ f \left(2^{-3/4}
\left| (p-q) e^{t_0} \right| \ e^{- x^+}\right) \delta {\cal U}_0
(p,q) , \\ [3mm] 
{\cal J}_{\rm out} (x^-) &=& 2^{-1/2} \int dp \ dq \ f \left(2^{-3/4}
\left| (p+q) e^{t_0} \right| \ e^{x^-}\right) \delta {\cal U}_0
(p,q) .
\eeq
These expressions for ${\cal J}_{\rm in}$ and ${\cal J}_{\rm out}$ do
not use any specific form of the profile of the fluctuation of the
fermi surface. Moreover, no specific parametrization of the fluctuation
of the fermi surface been used in eqns. (2.7) and (2.8). In practice,
however, we need to use some parametrization. In the following we will
use the familiar collective field parametrization, and to do so we
will need to assume a quadratic profile for $\delta {\cal U}_0 (p,q)$.
We emphasize that eqns. (2.7) and (2.8) may be used with any other
convenient parametrization. We will make a crucial use of this freedom
in Sec. 5. We also emphasize that the restriction to a quadratic
profile for $\delta {\cal U}_0 (p,q)$ does not restrict us to
quadratic profiles at all times, as in eqns. (2.1) and (2.2). For
sufficiently large fluctuations, in fact, the number of folds changes
(in the collective field parametrization) in time. It is, however,
clear that in our formalism there cannot be any physical effects
associated with such a change since the ${\cal J}_{\rm in}$ and
${\cal J}_{\rm out}$ in eqns. (2.7) and (2.8) depend only on the
nature of the profile of the initial distribution $\delta {\cal U}_0
(p,q)$\footnote{In particular, our formulation makes it clear that the
fold radiation discussed in [1] is a singularity associated with the
collective field parametrization. There cannot be any physical effect
associated with it}. 

To proceed further, we will use the independence of the right hand
sides of eqns. (2.7) and (2.8) from the parameter $t_0$ to choose a
convenient value for it. In fact, we will take $t_0 \rightarrow -
\infty$, so that $\delta {\cal U}_0 (p,q)$ indeed refers to the
initial configuration. Since fluctuations on both sides of the
potential are included (Fig. 3) in eqns. (2.7) and (2.8), we need two
independent massless fields, $\bar \eta^1_{+ \rm in}$ and
$\bar\eta^2_{+ \rm in}$, similar to the single field $\bar \eta_{+ \rm
in}$ which appeared in eqn. (2.1), to describe fluctuations on each of
the two sides of the potential. Actually, the following two
combinations of these fields appear more naturally in the formalism:
\beq
\phi (\tau) &\equiv& \left( \bar\eta^1_{+ \rm in} (\tau) +
\bar\eta^2_{+ \rm in} (\tau)\right)/ \sqrt{2} , \nonumber \\ [2mm]
\Delta (\tau) &\equiv& \left(\bar\eta^1_{+ \rm in} (\tau) -
\bar\eta^2_{+ \rm in} (\tau)\right)/ \sqrt{2} .
\eeq
In terms of these fields eqns. (2.7) and (2.8) may be rewritten as
\beq
{\cal J}_{\rm in} (x^+) &=& \int^{+\infty}_{-\infty} d \tau \ \phi
(\tau) \ f \left(\left|{\mu' \over 2}\right|^{1/2} \ e^{\tau -
x^+}\right) , \\ [3mm]
{\cal J}_{\rm out} (x^-) &=& {1 \over 2} \int^{+\infty}_{-\infty} d
\tau \bigg[ \int^{\phi(\tau)+\Delta (\tau)}_0 d \varepsilon \ f
\left(\left|{\mu' \over 2}\right|^{1/2} \left(1 - {\varepsilon \over
|\mu'|} \right) e^{-\tau + x^-}\right) \nonumber \\ [3mm]
&& ~~ + \int^{\phi (\tau) - \Delta (\tau)}_0 d \varepsilon \ f
\left(\left|{\mu' \over 2}\right|^{1/2} \left(1 - {\varepsilon \over
|\mu'|} \right) e^{-\tau + x^-} \right) \bigg] .
\eeq
The details of derivation of these equations may be found in ref.
[11]. Here $\mu' \equiv \sqrt{2} \mu$. 

There are two things that are immediately obvious from eqns. (2.10) and
(2.11). The first, as seen from eqn. (2.10), is that it is the
{\it sum} of 
the fluctuations on the two sides of the potential that maps onto the
tachyon of string theory. For $\Delta (\tau) = 0$ we recover the
tachyon scattering amplitudes of string theory is the backgrounds of
flat space, linear dilaton and linear tachyon.

The other thing that is clear from eqns. (2.10) and (2.11) is that
${\cal J}_{\rm out} (x^-)$ does not vanish when $\varphi (\tau) = 0$
(i.e. when the incoming tachyon field vanishes) but $\Delta (\tau)
\neq 0$. Classically the only way to interpret this is that for
$\Delta (\tau) \neq 0$ there is a new time-dependent tachyon
background present, over and above the linear tachyon background of
flat space. This new tachyon background vanishes as $t \rightarrow -
\infty$, and as $t \rightarrow + \infty$ it is given by
\beq
\left({\cal J}_{\rm out} (x^-)\right)_{\rm background} &=& {1 \over 2}
\int^{+\infty}_{-\infty} \ d \tau \bigg[ \int^{\Delta(\tau)}_0 d
\varepsilon \ f \left(\left|{\mu'\over 2}\right|^{1/2} \left(1 -
{\varepsilon \over |\mu'|}\right) e^{-\tau + x^-}\right) \nonumber \\
[3mm]
&&~~~~~~~~~+ \left(\Delta (\tau) \rightarrow - \Delta (\tau)\right)\bigg]
\eeq
Fortunately there is a nontrivial check we can perform to test this
interpretation since any background should be reflected
in the tachyon scattering amplitudes. In particular, the $1
\rightarrow 1$ tachyon bulk scattering should see the proposed
background in eqn. (2.12). Actually, as it turns out, the $1
\rightarrow 1$ tachyon bulk scattering amplitude for $\Delta (\tau)
\neq 0$, which is easily obtained from eqns. (2.10) and (2.11),
contains much more than just the amplitude for scattering off the
tachyon background identified in eqn. (2.12). A detailed perturbative
analysis (for small $\Delta$) at the first nontrivial order in
$\Delta$ has been carried out in ref. [11] for $1 \rightarrow 1$
tachyon bulk scattering and the leading term in the amplitude at early
times $(x^- \rightarrow - \infty)$ is compared with what is expected from
an analysis of $\beta$-function equations satisfied by tachyon,
dilaton and metric of 2-dimensional string theory. In addition to
the term expected from scattering off a tachyon background of the form
given in eqn. (2.12), the $1 \rightarrow 1$ bulk scattering amplitude
is seen to contain another term which would be expected if there were
a background metric given by the line element
\be
(ds)^2 = (1 - M \ e^{-4x}) (dt)^2 - (1 + M \ e^{-4x}) (dx)^2 ,
\ee
where the mass parameter $M$ equals the energy carried by the field
$\Delta (\tau)$, i.e.
\be
M = {1 \over 4\pi} \int^{+\infty}_{-\infty} d \tau \ \Delta^2 (\tau) .
\ee

We thus see that classically the space-time interpretation of a
nonzero value of $\Delta (\tau)$ is that the tachyon is propagating in
space-time background fields. In particular, we have identified the
metric perturbation modulus in terms of the microscopic fermion theory
variables. We also see that a consistent interpretation of space-time
gravitational physics energies in our formalism since the metric
couples to the total energy of the system, i.e. to the sum of energies
carried by the field $\phi (\tau)$, which gives the tachyon
fluctuation, and the field $\Delta (\tau)$, which gives the
backgrounds. It is this sum of energies that equals the total
Hamiltonian of the underlying fermion theory. By retaining
{\it both} sides of the potential we have been able to ``see''
the discrete state moduli of 2-dimensional string theory. It is,
therefore, only for this choice that {\it all} the tree-level
physics of 2-dimensional string theory can be extracted from the
matrix model.

\vskip 2em

\section{The linear tachyon background}

Actually there is one aspect of the tree-level physics of
2-dimensional string theory whose origin we have not yet explained in
the matrix model.  This is the existence of a linear tachyon
background in flat space in string theory [15].  To be sure this linear
tachyon background can indeed be seen in the matrix model by analysing
the $1 \rightarrow 1$ tachyon bulk scattering amplitude given by eqns.
(2.10) and (2.11) for $\Delta (\tau) = 0$ [10].  However, this does not
explain the origin of this background in the matrix model, especially
in view of the fact that all the other backgrounds arise from a
nonzero value of $\Delta (\tau)$. We would now like to show that the
linear tachyon background in flat space-time, in fact, arises from the
fermi vacuum.

To see this we go back go eqn. (2.4) and examine it more closely. The
form of this equation suggests that there might be a mapping of the
phase space density itself (and not just its fluctuations) to the
unshifted space-time tachyon field $T (x, t)$:
\beq
T (x,t) &=& e^{-2x} \ S (x,t) , \nonumber \\ [3mm]
S (x,t) &=& 2^{-1/2} \int dp \ dq \ f (2^{1/4} |q| e^{-x}) {\cal U}
(p, q, t) + O (e^{-2x})
\eeq
If this were the case then eqn. (2.4) would follow from eqn. (3.1) for
fluctuations around some ``classical'' fluid density configuration, which
in the present case is the fermi vacuum density ${\cal U}_F (p,q)$.
This ansatz would give rise to a time-independent tachyon background,
$T_F (x)$, corresponding to the fermi vacuum. In fact this turns out
to be precisely the linear tachyon background of string theory in flat
space. 

Let us verify the above statement. There is a slight subtility
encountered in doing this, namely the right hand side of eqn. (3.1) is
divergent for ${\cal U} = {\cal U}_F$. This divergence is associated
with the infinite number of fermions present in the double-scaled
theory and is a nonuniversal term. To get a finite result one needs to
subtract this term from the right hand side of eqn. (3.1) before
taking the double-scaling limit. One way to do so is to recognize that
the divergent piece does not depend on $\mu$ and so it may be got rid
of by taking a derivative of $S_F (x)$ with respect to $\mu$. This
gives
$$
\partial_\mu S_F (x) = 2^{-1/2} \int dp\:dq \ f (2^{1/4}|q| e^{-x})
\partial_\mu {\cal U}_F (p,q) + O (e^{-2x})
$$
Since an exact expression for ${\cal U}_F$ is known, the right hand
side of this equation can be calculated exactly, but we will here
confine ourselves to the semiclassical limit, $|\mu| \rightarrow
\infty$. In this limit ${\cal U}_F (p,q) = \theta (\mu - {1\over 2}
(p^2-q^2))$, and then we get
\be
\partial_\mu S_F (x) = {1 \over \sqrt{2\pi}} \left[2x + 4 \Gamma^\prime
(1) - ln \left({|\mu| \over \sqrt{\pi}}\right)\right] + O (e^{-x})
\ee
This shows that $T_F (x)$ has the form $(ax + b) e^{-2x}$ to leading
order in $e^{-x}$. The coefficients $a$ and $b$, which may be obtained
from eqn. (3.2), match exactly with those obtained from $1 \rightarrow
1$ tachyon bulk scattering amplitude in flat space.

We are now in a position to give a string theory interpretation to
{\it all} the states of the underlying microscopic fermi theory.
So far we have been treating the field $\Delta (\tau)$ classically.
However, in the quantum theory {\it both} $\varphi (\tau)$ and
$\Delta (\tau)$ are fluctuating quantum fields. The above results are
obtained by considering those fermi states in which there are a few
$\varphi$ excitations, but in which $\Delta$ has an expectation value,
with small quantum fluctuations.  A general state in the microscopic
theory, however, has both $\varphi$ and $\Delta$ excitations and both
the fields undergo quantum fluctuations. As we have seen, the former
appear in the string theory as tachyons. The later may now be seen to
give rise to fluctuating quantum fields corresponding to the metric
and higher tensor fields of 2-dimensional string theory! The
correspondence with the $W^\pm$ operators of [16] is that the $W^-$
map onto the states of the fermi theory and $W^+$ map onto the
$W$-infinity operators of the fermi theory [17].

\vskip 3em

\section{Black hole}

In this section we will identify some classical configurations of the
field $\Delta (\tau)$ which seem to have properties of the eternal
black hole of 2-dimensional string theory [18].

The Hamiltonian of the fermion theory, in terms of the asymptotic
``in'' and ``out'' fields, is given by
\beq
H &=& {1 \over 4\pi} \int^{+\infty}_{-\infty} d\tau \ [\varphi^2
(\tau) + \Delta^2 (\tau)] \nonumber \\ [3mm]
&=& {1 \over 4\pi} \int^{+\infty}_{-\infty} d\tau \ [\varphi^2_{\rm
out} (\tau) + \Delta^2_{\rm out} (\tau)] .
\eeq
We have dropped the subscript ``in'' in keeping with our earlier
usage. The ``out'' fields can be related to the ``in'' fields in
perturbation theory as expansions in string coupling ($\sim
|\mu|^{-1}$). We will be interested primarily in the situations in
which $\varphi (\tau) =0$. This corresponds to those configurations in
space-time in which the ``in'' tachyon field vanishes. In this case
the ``out'' fields in the microscopic fermion theory have the
following perturbative expansions in terms of the only nonvanishing
``in'' field $\Delta (\tau)$:
\beq
\varphi_{\rm out} (\tau) &=& - \left({1 \over 2 \sqrt{2} |\mu|}\right)
\partial_\tau \Delta^2 (\tau) + O (\Delta^4/|\mu|^2) , \\ [3mm]
\Delta_{\rm out} (\tau) &=& \Delta (\tau) + {1 \over 12 |\mu|^2}
\partial_\tau (\partial_\tau - 1) \Delta^3 (\tau) + O
(\Delta^5/|\mu|^4) .
\eeq
A check on these expansions is that the second equality of eqn. (4.1)
must be satisfied. This is indeed the case, order by order in the
string coupling, for the expansions for ``out'' fields
given in eqns. (4.2) and (4.3).

Consider now the class of configurations for which $\Delta (\tau)$ is
a slowly varying function of $\tau$. More precisely, let $L$ be the

scale over which $\Delta (\tau)$ varies by an appreciable fraction of
itself. In the asymptotic past the field $\varphi$ carries no energy
while the energy carried by the field $\Delta$ scales as $L$. In the
asymptotic future some of this energy is carried away by the field
$\varphi$. From eqns. (4.1)--(4.3) we see that, as a fraction of the
initial energy, this energy vanishes as $1/L^2$ for large $L$.
Combining this with our discussion in Sec. 2, we see that sufficiently
slowly varying configurations of $\Delta (\tau)$ correspond to
essentially {\it static metric backgrounds} in their space-time
interpretation. These backgrounds are characterized by a mass
parameter, given by the expression in eqn. (2.14), which takes
continuous real values. This is large for large $L$. Since the only
static metric solutions of the $\beta$-function equations of
2-dimensional string theory correspond to the eternal black hole, one
is led to suspect that the space-time manifestation of the
configurations we have identified here might be 
the eternal black hole.

To test the above hypothesis we will perform a scattering
``experiment''. Let us throw a beam of tachyons of
frequency $\omega$ at such a background. A classical black hole must
absorb a part of the incoming tachyon flux. For slowly varying $\Delta
(\tau)$, in the weak field approximation, $|\Delta (\tau)|/|\mu| \ll
1$, we can compute the $1 \rightarrow 1$ tachyon scattering amplitude
from the following relation which may be derived from eqns. (2.10) and
(2.11): 
\beq
{\tilde{\cal J}}_{\rm out} (\omega) &=& - \left({\sqrt{\pi} \over
|\mu|}\right)^{i\omega} \left({\Gamma (i\omega) \over \Gamma
(-i\omega)}\right)^2 \Re {\tilde{\cal J}}_{\rm in} (\omega) ,
\nonumber \\ [3mm]
\Re &=& \left[ 1 - \omega (\omega - i) {\Delta^2_0 \over 4|\mu|^2} + O
\left({\Delta_0^4 \over |\mu|^4}\right)\right] .
\eeq
Here $\Delta_0$ denotes the value of $\Delta (\tau)$ at $\tau = 0$ and
${\tilde{\cal J}}_{{\rm in}({\rm out})}$ is the Fourier transform of ${\cal
J}_{{\rm in}({\rm out})}$:
$$
{\tilde{\cal J}}_{{\rm in}({\rm out})} (\omega) \equiv \left({|\omega|
\over \pi}\right)^{1/2} \int^{+ \infty}_{-\infty} \ dx \ e^{i\omega x}
{\cal J}_{{\rm in ({\rm out})}} (x) .
$$
Note that in writing down eqn. (4.4) we have retained only the leading
term in a ${1 \over L}$ expansion. The leading result is independent
of $L$. So this result is applicable to backgrounds for which the mass
parameter $M \rightarrow \infty$. Since $|\Re|^2 < 1$, we see that the
space-time backgrounds under study are absorbing!

Let us summarize: We have identified a class of configurations in the
matrix model which have the space-time interpretation of being nearly
static (strictly so for $L \rightarrow \infty$) metric backgrounds
characterized by large continuous values $(\rightarrow \infty$ as $L
\rightarrow \infty)$ of the mass parameter. These metric backgrounds
are absorbing --- only a part of an incident beam of tachyons is
reflected back. Moreover, as argued at the beginning of this section,
nearly nothing (strictly so for $L \rightarrow \infty$) ever comes
out. It would be interesting to investigate whether one
could get at the exact space-time metric for these matrix model
backgrounds in the present formalism. 
There is one interesting question that the reader might ask at this
stage: Since the microscopic fermion theory is manifestly unitary,
what happens to the tachyons that are absorbed? Fortunately we can
answer this question rather easily. We must remember that in the
microscopic theory the field $\Delta$ is as much of a fluctuating
quantum field as $\varphi$ is. We must, therefore, also take the
fluctuations of $\Delta$ around its classical value into account. When
we do that we find the following relations between the ``in'' and
``out'' modes of the tachyon and the fluctuations of $\Delta$, which
we shall denote by $\delta$:
\beq
{\tilde{\cal J}}_{\rm out} (\omega) &=& - \left({\sqrt{\pi}\over
|\mu|}\right)^{{i\omega \over 2}} {\Gamma (i\omega) \over \Gamma
(-i\omega)} \ {\tilde\Re} \ \tilde\delta_{\rm in} (\omega) -
\left({\sqrt{\pi} \over |\mu|}\right)^{i\omega} \left({\Gamma
(i\omega) \over \Gamma(-i\omega)}\right)^2 \Re {\tilde{\cal J}}_{\rm
in} (\omega) , \nonumber \\ 
&& \\ [2mm]
{\tilde\delta_{\rm out}} (\omega) &=& - \left({\sqrt{\pi} \over
|\mu|}\right)^{{i\omega \over 2}} {\Gamma (i\omega) \over \Gamma
(-i\omega)} \ \tilde\Re \ {\tilde{\cal J}}_{\rm in} (\omega) + \Re
\tilde\delta_{\rm in} (\omega) .
\eeq
Here $\tilde\delta_{\rm in (out)} (\omega)$ is the Fourier transform of
$\delta (\tau)$:
$$
\tilde\delta_{\rm in (out)} (\omega) = 2\pi |\omega|^{-1/2}
\int^{+\infty}_{-\infty} d\tau \ e^{i\omega\tau} \delta_{\rm in (out)}
(\tau) .
$$
Also, $\Re$ is given by the expression in eqn. (4.4) and ${\tilde\Re}
= {\omega\Delta_0 \over \sqrt{\pi}|\mu|} + 0 \left({\Delta_0^3 \over
|\mu|^2}\right)$. Higher order terms in the ``in'' fields in these
equations have been dropped since we are interested only in the $1
\rightarrow 1$ scattering amplitudes. Eqn. (4.6) shows that an
incoming tachyon absorbed by the background metric is actually
converted into a $\delta$ excitation, producing an excited background
metric. Eqn. (4.5) shows that these excited background metric states
will decay into outgoing tachyons. The amplitudes for these two
processes are such that no probability is ever lost and the theory is
unitary and time-reversal invariant, as expected.

\vskip 2em

\section{Large fluctuations -- strong fields}

It has been argued by Polchinski [12] that a potential nonperturbative
inconsistency exists in the space-time interpretation of the matrix
model. His argument concerns large fluctuations of the fermi surface
which cross the asymptotes (Fig. 2) and is based on the conservation
of $W$-infinity charges and an identification of the weak field
expansion parameter (WFEP) which allows a perturbative treatment even
for large fluctuations. In this section we will argue that the
identification of the WFEP in [12] is not correct. We will argue that
for large fluctuations in the fermi theory perturbation expansion is
not valid in its string theory interpretation.

In the standard collective field parameterization of the fermi surface
(assumed to have a quadratic profile) the WFEP can be easily seen to
be $e^{-2\tau} \eta (\tau)/|\mu|$, where $\eta (\tau)$ is the
asymptotic ``in'' field that satisfies the equation of a free massless
particle. Here $\tau$ is large positive. This suggests that
perturbation theory would be valid even for $|\eta (\tau)|/|\mu| \gg
1$, provided we stay at sufficient large $\tau$ such that the
combination appearing in the WFEP is still small. On the string theory
side, the $\beta$-function equation of motion of the tachyon field,
${\cal J} (x,t)$, shows that the string theory WFEP is $e^{-2x} {\cal
J} (x,t)/|\mu|$. Thus large values of $|{\cal J} (x,t)|/|\mu|$ do not
necessarily invalidate perturbation expansion provided measurements 
are made at sufficiently large values of $x$. This would seem to fit in
with the observation of Polchinski made above in the matrix model. An
examination of the expression for ${\cal J}_{\rm out} (x^-)$ in eqn.
(2.11), however, shows that a perturbation expansion (assume $\Delta =
0$) is not valid unless we take $|\eta (\tau)|/|\mu| \ll 1$. What is
the source of this conflict?
 
There are two crucial points that need to be appreciated to understand
the source of the conflict pointed above. One is that large values of
${\cal J} (x,t)/|\mu|$ {\it do not} necessarily require large
values of $\eta (\tau)/|\mu|$. This is most clearly seen in eqns.
(2.7) and (2.8) which show that the magnitude of the tachyon field is
controlled not just by the {\it height} of the fluctuation but by
the phase space {\it area} occupied by it. Thus large values of
${\cal J}/|\mu|$ may be obtained from small but well spread out
fluctuations of the fermi surface!

The second point has to do with the use of collective field
parametrization of the profile of the fermi surface. We have stressed
earlier (in Sec. 2) that the physics of the space-time tachyon field
should not depend on any specific parametrization of the fermi surface.
This was explicitly incorporated in eqns. (2.7) and (2.8) which do not
depend on any specific parametrization. So let us use these equations
to understand better what is going on.

\begin{figure}[htb]
\centerline{\epsfxsize=3.5in\epsfbox{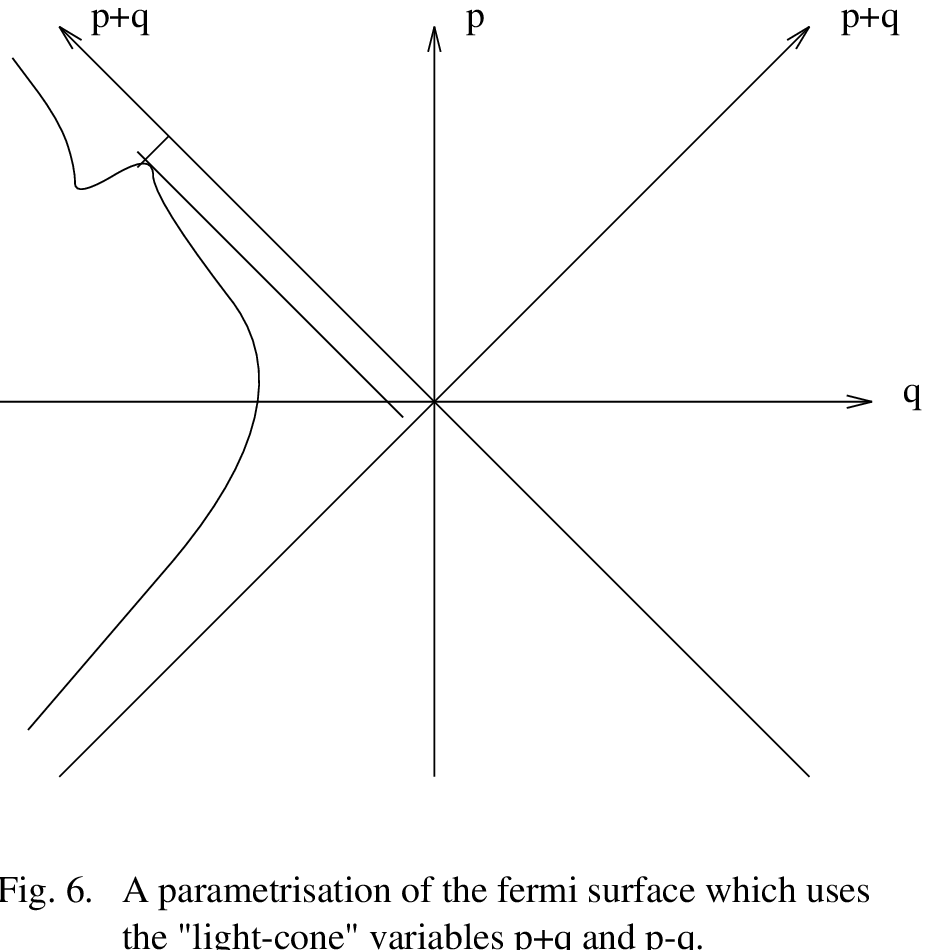}}
\end{figure}

Let us use an alternative parametrization of the fermi surface due to
O'Loughlin [19]. In this parametrization, say for the initial
configuration of the fermi surface, the distance from the asymptote
$p+q = 0$ is taken to be the field variable, $\psi (r,t)$, which
depends on the distance $r$ of the point under consideration from the
other asymptote, $p-q = 0$ (Fig. 6).
The WFEP in this case turns out to be
$\eta/|\mu|$, where $\eta$ is the field that satisfies the equation of
a free massless particle. This is quite different from the WFEP that we 
got using the collective field parametrization. However, 
the expressions for ${\cal J}_{\rm in}$ and
${\cal J}_{\rm out}$, as calculated from eqns. (2.7) and (2.8) using
this parametrization, work out to be mathematically
{\it identical} to those computed using the collective field
parametrization, thereby giving the same tachyon scattering 
amplitudes and hence the same space-time physics! This is an explicit 
verification of the statement that in our formalism the physics 
of the tachyon of string theory  depends only on the profile of 
the fluctuation of the fermi surface and not on any specific 
parametrisation used. It is also now clear that the relevant WFEP 
in the fermi theory, which gives the weak field expansion in 
string theory, is best read off from eqns. (2.7) and (2.8). 

We conclude that the WFEP in the fermi theory is $\eta/|\mu|$ so that
weak fields in the matrix model are identical to small fluctuations.
Strong fields correspond to large fluctuations for which perturbation
theory breaks down. In the fermi theory this is accompanied by some of 
the fluid crossing over to the other side, perhaps signalling some new
physics. We emphasize that this conclusion is {\it not} in conflict
with the fact that string perturbation expansion is valid even for 
large values
of ${\cal J}/|\mu|$ since, as explained earlier, small values of
$\eta/|\mu|$ already incorporate the possibility of having large
values of ${\cal J}/|\mu|$.

\vskip 2em

\section{Concluding remarks}

Extracting the space-time physics of 2-dimensional string theory from
the $c=1$ matrix model has been a long, hard and often frustrating
enterprise. It is, therefore, a matter of some satisfaction that we
now have a complete mapping from the matrix model to the tree-level
physics of string theory, including the discrete state moduli. This
correspondence associates flat space-time, linear dilaton and linear
tachyon background with the fermi sea of the underlying fermi
system. Excited states in the fermi theory are in general associated
with tachyons propagating in nontrivial backgrounds of the space-time
metric and higher tensor fields.  Moreover, as we have argued, there
does not seem to be any potential inconsistency in taking the matrix
model to define nonperturbative 2-dimensional string theory. What the
matrix model might teach us about the nonperturbative structure of
string theory remains to be seen, but there are at least two important
lessons that have already come through. One is about the nature of
microscopic degrees of freedom --- these are nothing like strings at
all!  The other is about the subtle and totally unexpected way in
which space-time gravitational physics emerges from the microscopic
theory which does not seem to ``know'' anything about gravitational
physics. It is interesting that the currently popular proposal for a
microscopic formulation of nonperturbative superstring theory in
10-dimensions [20] shares these two features with the $c=1$ matrix
model.

\vskip 2em

\noindent{\bf Acknowledgements}

It is a great pleasure to acknowledge numerous useful discussions with
G. Mandal and S.R. Wadia on all aspects of the subject treated in this
work.

\newpage

\noindent\underbar{\bf References} 
\begin{enumerate}
\item{} J. Polchinski, ``What is String Theory?'', Proceedings of
the 1994 Les Hauches Summer School, edited by F. David, P. Ginsparg
and J. Zinn-Justin (Elsevier, Amsterdam, 1996) p. 287, hep-th/9411028.
\item{} D.J. Gross and A. Migdal, {\it Phys. Rev. Lett.} {\bf 64}
(1990) 127; M. Douglas and S. Shenker, {\it Nucl. Phys.} {\bf B 335}
(1990) 635; E. Brezin and V.A. Kazakov, {\it Phys. Lett.} {\bf 236}
(1990) 144.
\item{} E. Brezin, V.A. Kazakov and A.B. Zamolodchikov, {\it Nucl.
Phys.} {\bf B 333} (1990) 673; P. Ginsparg and J. Zinn-Justin, {\it
Phys. Lett.} {\bf B 240} (1990) 333; G. Parisi, {\it Phys. Lett.} {\bf
B 238} (1990) 209; D.J. Gross and N. Miljkovic, {\it Phys. Lett.} {\bf
B 238} (1990) 217.
\item{} S.R. Das and A. Jevicki, {\it Mod. Phys. Lett.} {\bf A 5}
(1990) 1639.
\item{} J. Polchinski, {\it Nucl. Phys.} {\bf B 362} (1991) 125.
\item{} A.M. Sengupta and S.R. Wadia, {\it Int. J. Mod. Phys.}
{\bf A 6} (1991) 1961.
\item{} P. Di Francesco and D. Kutasov, {\it Phys. Lett.} {\bf B
261} (1991) 385 and {\it Nucl. Phys.} {\bf B 375} (1992) 119.
\item{} A.M. Polyakov, {\it Mod. Phys. Lett.} {\bf A6} (1991) 635.
\item{} D.J. Gross and I. Klebanov, {\it Nucl. Phys.} {\bf B 359}
(1991) 3.
\item{} M. Natsuume and J. Polchinski, {\it Nucl. Phys.} {\bf B
424} (1994) 137.
\item{} A. Dhar, G. Mandal and S.R. Wadia, {\it Nucl. Phys.} {\bf
B 454} (1995) 541.
\item{} J. Polchinski, {\it Phys. Rev. Lett.} {\bf 74} (1995) 63.
\item{} A. Dhar, G. Mandal and S.R. Wadia, {\it Nucl. Phys.} {\bf
B 451} (1995) 507.
\item{} A. Dhar, G. Mandal and S.R. Wadia, {\it Mod. Phys. Lett.}
{\bf A 7} (1992) 3129 and {\it Mod. Phys. Lett.} {\bf A 8} (1993) 3557.
\item{} J. Polchinski, {\it Nucl. Phys.} {\bf B 346} (1990) 253.
\item{} E. Wilten, {\it Nucl. Phys.} {\bf B 373} (1992) 187.
\item{} J. Avan and A. Jevicki, {\it Phys. Lett.} {\bf B 266}
(1991) 35; G. Moore and N. Seiberg, {\it Int. J. Mod. Phys.} {\bf A 7}
(1992) 2601; S.R. Das, A. Dhar, G. Mandal and S.R. Wadia, {\it Int. J.
Mod. Phys.} {\bf A 7} (1992) 5165; D. Minic, J. Polchinski and Z.
Yang, {\it Nucl. Phys.} {\bf B 369} (1992) 324.
\item{} G. Mandal, A. Sengupta and S.R. Wadia, {\it Mod. Phys.
Lett.} {\bf A 6} (1991) 1685; E. Witten, {\it Phys. Rev.} {\bf D 44}
(1991) 314.
\item{} M. O'Loughlin, {\it Nucl. Phys.} {\bf B 449} (1995) 80.
\item{} T. Banks, W. Fishler, S. Shenker and L. Susskind, ``M
Theory as a Matrix Model: A Conjecture'', hep-th/9610043.
\end{enumerate}
\end{document}